\documentclass[superscriptaddress,twocolumn,english,floatfix,aps,pra,amsmath,amssymb,longbibliography]
{revtex4-2}
\usepackage[T1]{fontenc}
\usepackage[utf8]{inputenc}
\usepackage{babel}
\usepackage{comment}
\usepackage{color}
\usepackage{times}
\usepackage{epsfig}
\usepackage{subfigure}
\usepackage{amsmath}
\usepackage{amssymb}
\usepackage{graphicx}
\usepackage[colorlinks=true]{hyperref}
\hypersetup{citecolor=blue,linkcolor=blue,urlcolor=blue}

\begin{document}

\title{Sign inversion in the lateral van der Waals force between an anisotropic particle and a plane with a hemispherical protuberance: an exact calculation}

\author{Lucas Queiroz} 
\email{lucas.silva@icen.ufpa.br}
\affiliation{Faculdade de F\'{i}sica, Universidade Federal do Par\'{a}, 66075-110, Bel\'{e}m, Par\'{a}, Brazil}

\author{Edson C. M. Nogueira}
\email{edson.moraes.nogueira@icen.ufpa.br}
\affiliation{Faculdade de F\'{i}sica, Universidade Federal do Par\'{a}, 66075-110, Bel\'{e}m, Par\'{a}, Brazil}

\author{Danilo T. Alves}
\email{danilo@ufpa.br}
\affiliation{Faculdade de F\'{i}sica, Universidade Federal do Par\'{a}, 66075-110, Bel\'{e}m, Par\'{a}, Brazil}
\affiliation{Centro de F\'{i}sica, Universidade do Minho, P-4710-057, Braga, Portugal}

\date{\today}

\begin{abstract}
%
We investigate the lateral van der Waals (vdW) force between an anisotropic polarizable particle and a perfectly conducting plane with a hemispherical protuberance with radius $R$.
We predict, via an exact calculation, a sign inversion in the lateral vdW force, in the sense that, instead of pointing to the protuberance, in certain situations this force points to the opposite direction.
In the literature, predictions of sign inversions in the lateral vdW force were based on perturbative solutions, valid when the height of the protuberance is very small when compared to the distance $z_0$ between the particle and the plane.
Here, taking into account exact formulas, we investigate how such nontrivial geometric effect depends on the ratio $R/z_0$, and how the particle orientation and anisotropy affect this sign inversion.
\end{abstract}

\maketitle

\section{Introduction}

Investigations of the Casimir-Polder--van der Waals (CP-vdW) \cite{Casimir-Polder-PhysRev-1948} interaction considering anisotropic particles generally involves nontrivial behaviors.
Among them is the prediction of a repulsive CP-vdW force \cite{Levin-PRL-2010, Eberlein-PRA-2011, Buhmann-IJMPA-2016, Abrantes-PRA-2018, Venkataram-PRA-2020, Marchetta-PRA-2021} and a torque on a particle \cite{Bimonte-PRD-2015, Thiyam-PRA-2015, Gangaraj-PRB-2018, Antezza-PRB-2020}.
Beyond this, when considering the lateral CP-vdW force that acts on a particle interacting with a corrugated surface, nontrivial behaviors of this force are predicted when calculations are made beyond the proximity force approximation (PFA).
As an example, it was predicted the presence of regimes in the behavior of the lateral vdW force when an anisotropic polarizable particle interacts with a periodic corrugated surface \cite{Nogueira-PRA-2021,Queiroz-PRA-2021}.
In addition, recently it was predicted that,
for an anisotropic particle interacting with a conducting plane with a protuberance,
a sign inversion in the lateral vdW force can occur, in the sense that, instead of pointing to the protuberance, in certain situations the force points to the opposite direction \cite{Nogueira-PRA-2022}.

The prediction discussed in Ref. \cite{Nogueira-PRA-2022} was based on the application of a formula proposed by Eberlein and Zietal in Ref. \cite{Eberlein-PRA-2007} to compute the vdW interaction between a neutral polarizable particle and a perfectly conducting surface of arbitrary shape.
This formula is written in terms of the homogeneous part of the Green's function of the Laplacian operator,
where it is stored the information about the geometry of the surface.
In Ref. \cite{Nogueira-PRA-2022}, it was considered a perturbative analytical solution for the Green's function for the case of a perfectly conducting plane with a protuberance and, using it in the Eberlein-Zietal formula, it was predicted the mentioned sign inversion in the lateral vdW force up to first perturbative order in the ratio $a/z_0\ll 1$ ($z_0$ is the distance from the particle to the plane, and $a$ is the height of the protuberance). 

In Ref. \cite{Souza-AJP-2013}, the Eberlein-Zietal formula was combined with the well known solution
for the Green's function for the problem of a perfectly conducting plane with a hemispherical protuberance with radius $R$ \cite{Schwinger-Electrodynamics-1998}, and it was investigated the curvature effects of the hemisphere on the vdW force normal to the plane.
In the present paper, we use the same calculation technique but
focus our attention specifically on the lateral vdW force.
We confirm, now via an exact calculation, the sign inversion previously predicted in Ref. \cite{Nogueira-PRA-2022} in the context of a perturbative approach. We show how such nontrivial geometric effects are regulated by the ratio $R/z_0$, and how the existence of these effects depend on the particle orientation and its anisotropy.

The paper is organized as follows. 
In Sec. \ref{sec-brief-discussion}, we make a brief review of the perturbative predictions made in Ref. \cite{Nogueira-PRA-2022}.
In Sec. \ref{sec-exact-calculation}, we present the exact calculation made to write the formulas for interaction between a neutral polarizable particle and a conducting plane with a hemisphere.
In Sec. \ref{sec-final-remarks}, we present our final remarks.

\section{A brief review of the perturbative calculation}
\label{sec-brief-discussion}

\begin{figure}[h]
\centering
\epsfig{file=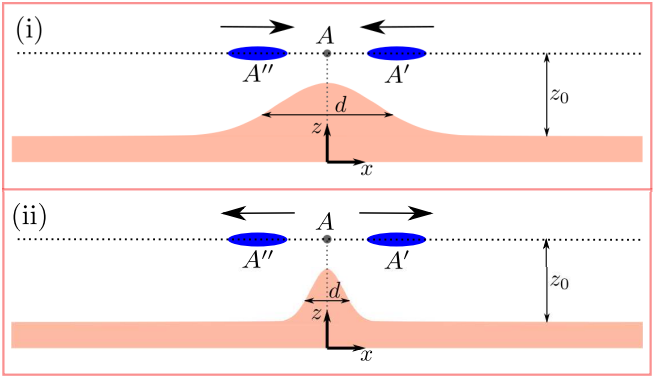,  width=1 \linewidth}
\caption{
Illustration of a neutral anisotropic polarizable particle (elliptic figures), kept constrained to move on the plane $z=z_0$ (horizontal dotted lines), interacting with a perfectly conducting plane with a single slight Gaussian protuberance with characteristic width $d$.
Due to the presence of the protuberance, the particle feels a lateral force (arrows).
In (i), the particle at $A^\prime$ (or $A^{\prime\prime}$) feels a lateral force that takes it back to $A$; 
in (ii), for a decreased ratio $d/z_0$, the lateral force moves the particle away from $A$.
This sign inversion in the lateral force was predicted in Ref. \cite{Nogueira-PRA-2022}.
}
\label{fig:gaussiana}
\end{figure}

In Ref. \cite{Nogueira-PRA-2022}, it was considered the vdW interaction between a plane surface $(z=0)$ with a single slight protuberance, and a neutral polarizable particle located at $\textbf{r}_0$, and kept constrained to move on a plane $z = z_0>0$ (Fig. \ref{fig:gaussiana}). 
To investigate this interaction, it was took as basis the analytical perturbative formula presented in Ref. \cite{Nogueira-PRA-2021}, related to the vdW interaction between the particle and a grounded conducting corrugated surface described by $z=h(\mathbf{r}_{\parallel})$, with $h({\bf r}_{||})$ describing a general small modification $[\text{max}|h({\bf r}_{||})|=a\ll z_0]$ of a planar surface at $z=0$ (in Ref. \cite{Nogueira-PRA-2022}, this modification is represented by a single protuberance on the surface).
The vdW interaction energy $U_\text{vdW}$, between the particle and the corrugated surface, was written as $U_\text{vdW}\approx U^{(0)}_{\text{vdW}} + U^{(1)}_{\text{vdW}}$, where the first term $U^{(0)}_{\text{vdW}}$ is the vdW potential for the case of a grounded conducting plane \cite{Lennard-Jones-TransFarSoc-1932}, whereas $U^{(1)}_{\text{vdW}}$ is the first perturbative correction of $U^{(0)}_{\text{vdW}}$ due to the presence of the corrugation, and is given by Eqs. (1)-(3) in Ref. \cite{Nogueira-PRA-2022}.
This approach requires no demand on the smoothness of the rough surface, so that the results are valid beyond the PFA.

The main situation discussed in Ref. \cite{Nogueira-PRA-2022} was the case of a Gaussian protuberance of height $a$ and width $d$, given by $h(\mathbf{r}_{\parallel})= a\exp[{-\left(|\mathbf{r}_{\parallel}|/d\right)^2}]$.
It was shown that when the PFA is considered ($d/z_0 \to \infty$), the lateral vdW force always leads the particle towards the peak of the protuberance.
On the other hand, for a generic value of $d/z_0$, nontrivial geometric effects arise when considering an anisotropic polarizable particle and manipulating the ratio $d/z_0$.
In this situation, the main prediction is that as $d/z_0$ decreases, the minimum value of the vdW energy can no longer coincide with the peak of the protuberance, which means that, instead of being attracted to the peak, the particle can be moved away from it (see Fig. \ref{fig:gaussiana}).
It was also shown in Ref. \cite{Nogueira-PRA-2022} that such a sign inversion in the lateral vdW force is affected by the particle orientation and its anisotropy, so that, depending on them, the mentioned effects can be suppressed.

As mentioned, in Ref. \cite{Nogueira-PRA-2022} these nontrivial behaviors of the lateral vdW force were predicted by means of a perturbative approach with $a/z_0 \ll 1$, and investigating them using only the first perturbative correction $U^{(1)}_{\text{vdW}}$.
Differently from this, in the next section, we study such nontrivial effects by means of an exact calculation, considering a perfectly conducting plane with a hemispherical protuberance.

\section{Exact calculation}
\label{sec-exact-calculation}

Let us start considering a point charge $Q$, located at the position ${\bf r}^{\prime}={\bf r}_{||}^{\prime}+z^\prime\hat{{\bf z}}$ (with $z^\prime>0$ and ${\bf r}_{||}^\prime=x^\prime\hat{{\bf x}}+y^\prime\hat{{\bf y}}$), interacting with a conducting plane surface $(z=0)$ that has a protuberance with the shape of a hemisphere of radius $R$, as shown in Fig. \ref{fig:calota-3D}.
The potential $\Phi\left({\bf r}\right)$, related to the Poisson's equation of this problem, can be obtained exactly by means of the image method.
For this case, one can obtain that only three image charges are necessary to obtain $\Phi\left({\bf r}\right)=0$ at the plane $z=0$ and the hemisphere \cite{Schwinger-Electrodynamics-1998,Souza-AJP-2013}.
In this way, one can obtain that the potential is given by 
\begin{equation}
\Phi\left({\bf r}\right)=\frac{Q}{\left|{\bf r}-{\bf r}^{\prime}\right|}+\frac{\tilde{Q}}{\left|{\bf r}-\tilde{{\bf r}}^{\prime}\right|}+\frac{\overline{Q}}{\left|{\bf r}-\overline{{\bf r}}^{\prime}\right|}+\frac{\tilde{\overline{Q}}}{|{\bf r}-\tilde{\overline{{\bf r}}}\,^{\prime}|},
\label{eq:phi-exato}
\end{equation}
where $\tilde{{\bf r}}^{\prime}$, $\overline{{\bf r}}^{\prime}$ and $\tilde{\overline{{\bf r}}}\,^{\prime}$ are the position vectors of the image charges $\tilde{Q}$, $\overline{Q}$ and $\tilde{\overline{Q}}$, respectively [see Fig. \ref{fig:calota-2D}] \cite{Schwinger-Electrodynamics-1998,Souza-AJP-2013}.
By writing these position vectors in cylindrical coordinates $(\rho,\phi,z)$,
one obtains \cite{Souza-AJP-2013}
\begin{widetext}
\begin{align}
\left|{\bf r}-{\bf r}^{\prime}\right| & =\sqrt{\rho^{2}+\rho^{\prime2}-2\rho\rho^{\prime}\cos\left(\phi-\phi^{\prime}\right)+\left(z-z^{\prime}\right)^{2}},\\
\left|{\bf r}-\tilde{{\bf r}}^{\prime}\right| & =\sqrt{\rho^{2}+\rho^{\prime2}-2\rho\rho^{\prime}\cos\left(\phi-\phi^{\prime}\right)+\left(z+z^{\prime}\right)^{2}},\\
\left|{\bf r}-\overline{{\bf r}}^{\prime}\right| & =\frac{1}{\left(\rho^{\prime2}+z^{\prime2}\right)}\sqrt{\rho^{2}\left(\rho^{\prime2}+z^{\prime2}\right)^{2}+R^{4}\rho^{\prime2}-2R^{2}\left(\rho^{\prime2}+z^{\prime2}\right)\rho\rho^{\prime}\cos\left(\phi-\phi^{\prime}\right)+\left[z\left(\rho^{\prime2}+z^{\prime2}\right)-R^{2}z^{\prime}\right]^{2}},\\
|{\bf r}-\tilde{\overline{{\bf r}}}\,^{\prime}| & =\frac{1}{\left(\rho^{\prime2}+z^{\prime2}\right)}\sqrt{\rho^{2}\left(\rho^{\prime2}+z^{\prime2}\right)^{2}+R^{4}\rho^{\prime2}-2R^{2}\left(\rho^{\prime2}+z^{\prime2}\right)\rho\rho^{\prime}\cos\left(\phi-\phi^{\prime}\right)+\left[z\left(\rho^{\prime2}+z^{\prime2}\right)+R^{2}z^{\prime}\right]^{2}},
\end{align}
and the charges $\tilde{Q}$, $\overline{Q}$ and $\tilde{\overline{Q}}$, in terms of $Q$, as
$\tilde{Q}=-Q$, $\overline{Q}=-{QR}/{\sqrt{\rho^{\prime2}+z^{\prime2}}}$,
and $\tilde{\overline{Q}}={QR}/{\sqrt{\rho^{\prime2}+z^{\prime2}}}$ \cite{Souza-AJP-2013}.
Since $\Phi\left({\bf r},{\bf r}^{\prime}\right) = Q G\left({\bf r},{\bf r}^{\prime}\right)$,
the Green's function $G\left({\bf r},{\bf r}^{\prime}\right)$ is given by
\begin{eqnarray}
G\left({\bf r},{\bf r}^{\prime}\right)=\frac{1}{\left|{\bf r}-{\bf r}^{\prime}\right|}-\frac{1}{\left|{\bf r}-\tilde{{\bf r}}^{\prime}\right|} -\frac{R}{\sqrt{\rho^{\prime2}+z^{\prime2}}}\left(\frac{1}{\left|{\bf r}-\overline{{\bf r}}^{\prime}\right|}-\frac{1}{|{\bf r}-\tilde{\overline{{\bf r}}}\,^{\prime}|}\right).
\label{eq:G-exato}
\end{eqnarray}
\begin{figure}[h]
\centering  
\subfigure[\label{fig:calota-3D}]{\epsfig{file=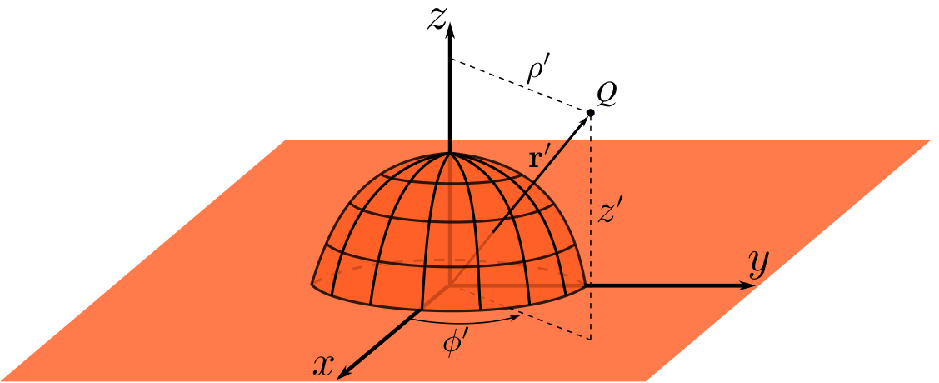, width=0.5 \linewidth}}
\hspace{1cm}
\subfigure[\label{fig:calota-2D}]{\epsfig{file=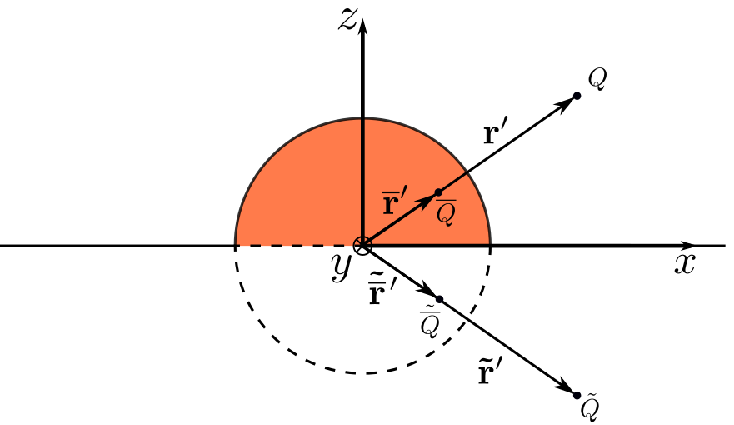, width=0.43 \linewidth}}
\caption{ 
(a) Illustration of a charge $Q$ located at ${\bf r}^\prime={\bf r}_{\parallel}^\prime+z^\prime\hat{{\bf z}}$ (with $z^\prime>0$), interacting with a conducting plane surface $(z=0)$ that has a protuberance with the shape of a hemisphere of radius $R$.
(b) Illustration of the charge $Q$ and its three image charges $\tilde{Q}$, $\overline{Q}$ and $\tilde{\overline{Q}}$, which are located at  $\tilde{{\bf r}}^{\prime}$, $\overline{{\bf r}}^{\prime}$ and $\tilde{\overline{{\bf r}}}\,^{\prime}$, respectively.
}
\label{fig:calota}
\end{figure}
\end{widetext}

Now that we have the Green's function related to a point charge in the presence of a conducting plane with a hemisphere, we can compute the vdW interaction $U_{\text{vdW}}$ between a polarizable particle located at $\textbf{r}_0$ and this surface, by using the formula proposed by Eberlein and Zietal in Ref. \cite{Eberlein-PRA-2007}, which is given by
\begin{equation}
U_{\text{vdW}}(\mathbf{r}_0) = \frac{1}{8\pi\epsilon_{0}}\displaystyle\sum_{i,j}\langle \hat{d}_i \hat{d}_j\rangle\boldsymbol{ \nabla}_i\boldsymbol{ \nabla}_j' \left.G_H(\mathbf{r},\mathbf{r'})\right|_{\mathbf{r}=\mathbf{r'}=\mathbf{r}_0},	\label{eq:Eberlein_Zietalquantum}
\end{equation}
where 
\begin{equation}
G_H\left(\textbf{r},\textbf{r}^{\prime}\right) = G\left(\textbf{r},\textbf{r}^{\prime}\right) - \frac{1}{|\textbf{r}-\textbf{r}^{\prime}|},
\label{eq:GH-exato}
\end{equation} 
and $\hat{d}_i$ are the components of the dipole moment operator.
Thus, using Eq. \eqref{eq:G-exato} in Eq. \eqref{eq:GH-exato}, and substituting in Eq. \eqref{eq:Eberlein_Zietalquantum}, one obtains that $U_\text{vdW} = U^{(0)}_{\text{vdW}} + U^{(\text{h})}_{\text{vdW}}$ \cite{Souza-AJP-2013}.
The first term $U^{(0)}_{\text{vdW}}$ is the vdW potential for the case of a grounded conducting plane \cite{Lennard-Jones-TransFarSoc-1932}, and is given by
\begin{equation}
U_{\text{vdW}}^{(0)}=-\frac{1}{64\pi\epsilon_{0}z_{0}^{3}}\left[\langle \hat{d}_{\rho}^{2}\rangle +\langle \hat{d}_{\phi}^{2}\rangle +2\langle \hat{d}_{z}^{2}\rangle \right].
\label{eq:U-0}
\end{equation}
The second term, $U^{(\text{h})}_{\text{vdW}}$, is the term of the energy that arises due to the presence of the hemisphere, and can be written as
\begin{widetext}
\begin{equation}
U_{\text{vdW}}^{(\text{h})}=-\frac{1}{64\pi\epsilon_{0}z_{0}^{3}}\left[\langle \hat{d}_{\rho}^{2}\rangle {\cal R}_{\rho\rho}+\langle \hat{d}_{\phi}^{2}\rangle {\cal R}_{\phi\phi}+\langle \hat{d}_{z}^{2}\rangle {\cal R}_{zz}+\langle \hat{d}_{\rho}\hat{d}_{z}\rangle {\cal R}_{\rho z}\right],
\label{eq:U-calota}
\end{equation}
where
\begin{align}
{\cal R}_{\rho\rho} & =-8\overline{R}\left\{ \frac{\overline{R}^{2}+\overline{\rho}_{0}^{2}}{\left(\overline{R}^{2}-\overline{\rho}_{0}^{2}-1\right)^{3}}+\frac{\overline{R}^{2}\left[\left(\overline{R}^{2}+1\right)^{2}-\overline{\rho}_{0}^{2}\left(\overline{R}^{2}+\overline{\rho}_{0}^{2}+8\right)\right]+\overline{\rho}_{0}^{2}\left(1+\overline{\rho}_{0}^{2}\right)^{2}}{\left[\overline{R}^{4}+2\overline{R}^{2}\left(1-\overline{\rho}_{0}^{2}\right)+\left(1+\overline{\rho}_{0}^{2}\right){}^{2}\right]^{5/2}}\right\} , \nonumber \\
{\cal R}_{\phi\phi} & =-8\overline{R}^{3}\left\{ \frac{1}{\left(\overline{R}^{2}-\overline{\rho}_{0}^{2}-1\right)^{3}}+\frac{1}{\left[\overline{R}^{4}+2\overline{R}^{2}\left(1-\overline{\rho}_{0}^{2}\right)+\left(1+\overline{\rho}_{0}^{2}\right)^{2}\right]^{3/2}}\right\} , \nonumber \\
{\cal R}_{zz} & =-8\overline{R}\left\{ \frac{\overline{R}^{2}+1}{\left(\overline{R}^{2}-\overline{\rho}_{0}^{2}-1\right)^{3}}-\frac{\overline{R}^{2}\left[\left(\overline{R}^{2}-\overline{\rho}_{0}^{2}\right)^{2}+\left(\overline{R}^{2}-1-8\overline{\rho}_{0}^{2}\right)\right]-\left(1+\overline{\rho}_{0}^{2}\right)^{2}}{\left[\overline{R}^{4}+2\overline{R}^{2}\left(1-\overline{\rho}_{0}^{2}\right)+\left(1+\overline{\rho}_{0}^{2}\right)^{2}\right]^{5/2}}\right\} ,\label{eq:rij-calota} \\
{\cal R}_{\rho z} & =-16\overline{R}\overline{\rho}_{0}\left\{ \frac{1}{\left(\overline{R}^{2}-\overline{\rho}_{0}^{2}-1\right)^{3}}-\frac{5\overline{R}^{4}+4\overline{R}^{2}\left(1-\overline{\rho}_{0}^{2}\right)-\left(1+\overline{\rho}_{0}^{2}\right)^{2}}{\left[\overline{R}^{4}+2\overline{R}^{2}\left(1-\overline{\rho}_{0}^{2}\right)+\left(1+\overline{\rho}_{0}^{2}\right)^{2}\right]^{5/2}}\right\} , \nonumber 
\end{align}
\end{widetext}
with $\overline{R}=R/z_0$ and $\overline{\rho}_0=\rho_0/z_0$.
Note that a lateral vdW force that acts on the particle arises due to the dependence of $U^{(\text{h})}_{\text{vdW}}$ on the variable $\rho_{0}$.
Since we are interested only in the behavior of this force, hereafter we focus our attention only on $U^{(\text{h})}_{\text{vdW}}$.
We remark that the behavior of the lateral force just depends on the ratios $R/z_0$ and $\rho_0/z_0$, and, thus, we investigate how the variation of $R/z_0$ affects the behavior of $U^{(\text{h})}_{\text{vdW}}$ in relation to $\rho_0/z_0$.
Besides this, one can note that for $R \to 0$, $U^{(\text{h})}_{\text{vdW}}$ vanishes, and we recover the energy interaction for the case of a grounded conducting plane \cite{Souza-AJP-2013}.

In Ref. \cite{Nogueira-PRA-2022}, the geometric effects due to the presence of the protuberance are regulated by the ratio between the characteristic width $d$ of the protuberance and the distance $z_0$ from the particle to the plane.
In a similar way, in the situation discussed here, the geometric effects due to the presence of the hemisphere are regulated by the ratio $R/z_0$.
In this way, we are going to investigate how the variation of this ratio affects the behavior of $U^{(\text{h})}_{\text{vdW}}$.
We remark that, unlike the perturbative approach discussed in Ref. \cite{Nogueira-PRA-2022}, by manipulating the width of the hemisphere we alter its size as a whole, and therefore the geometric effects predicted here are related to a change in the entire size of the protuberance, not just its width.

\begin{figure}[h]
\centering  
\subfigure[\label{fig:particula-cilindricas-3d}]{\epsfig{file=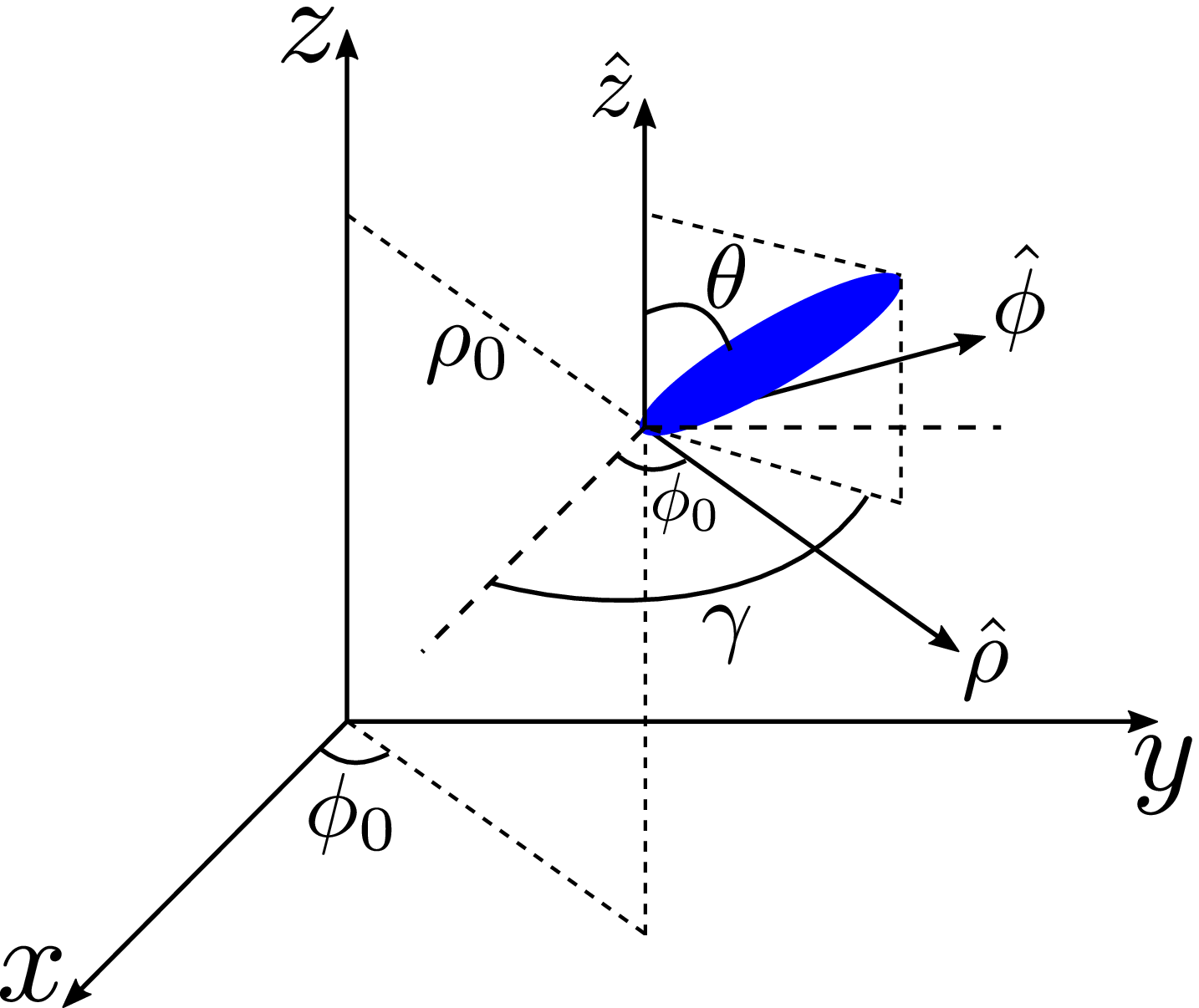, width=0.48 \linewidth}}
\hspace{1cm}
\subfigure[\label{fig:particula-cilindricas-2d}]{\epsfig{file=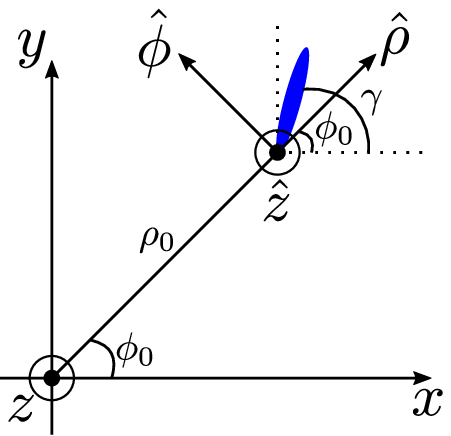, width=0.38 \linewidth}}
\caption{ 
Illustration of a polarizable particle characterized by Eq. \eqref{eq:didj} with a general orientation described by the angles $\theta$ and $\gamma$.
In (a), it is shown a 3D view, whereas in (b), a 2D one.
}
\label{fig:particula-cilindricas}
\end{figure}

\begin{figure}[h]
\centering  
\subfigure[\label{fig:energia-xy-r06}]{\epsfig{file=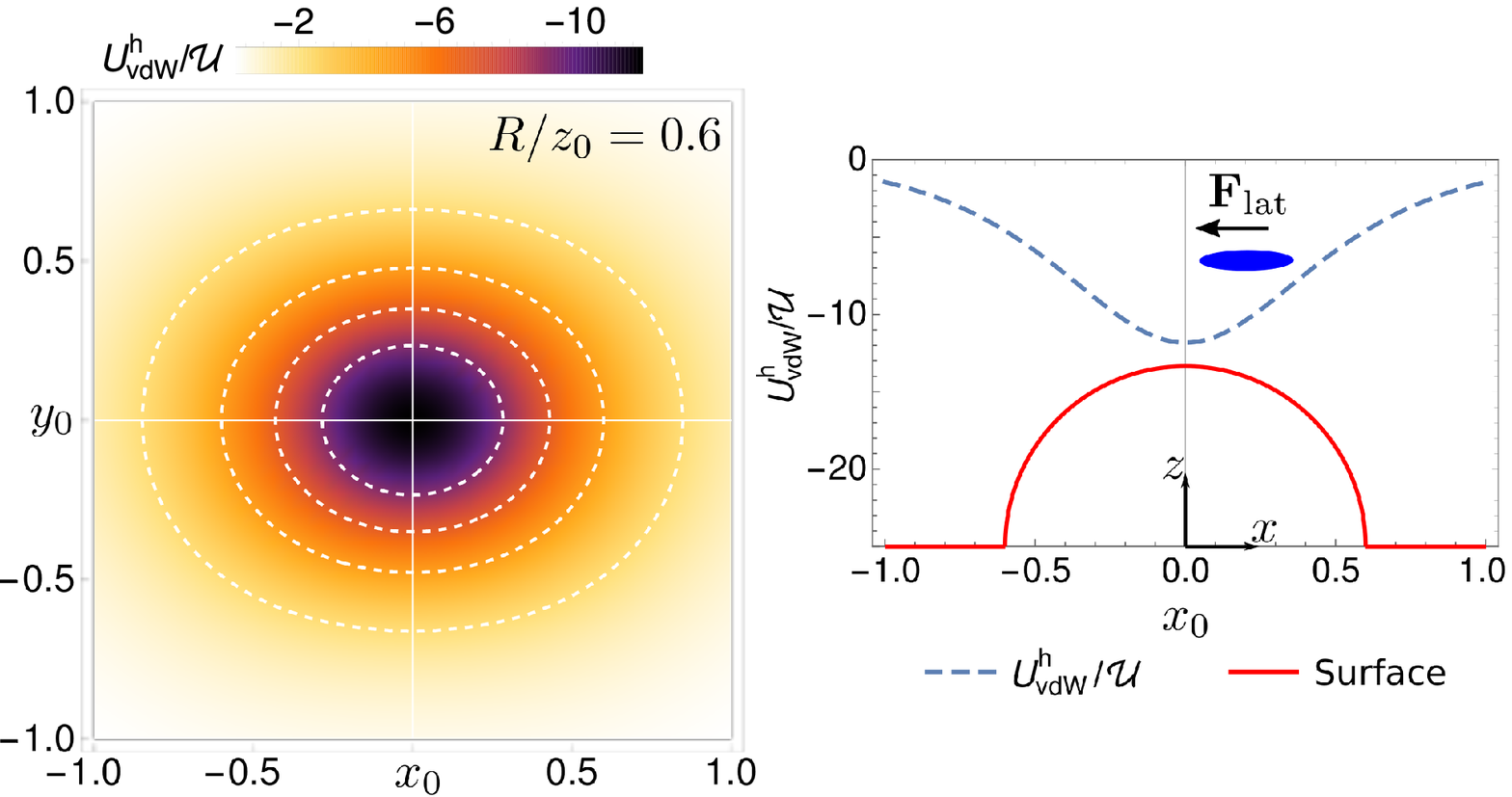, width=1 \linewidth}}
\hspace{4mm}
\subfigure[\label{fig:energia-xy-r02}]{\epsfig{file=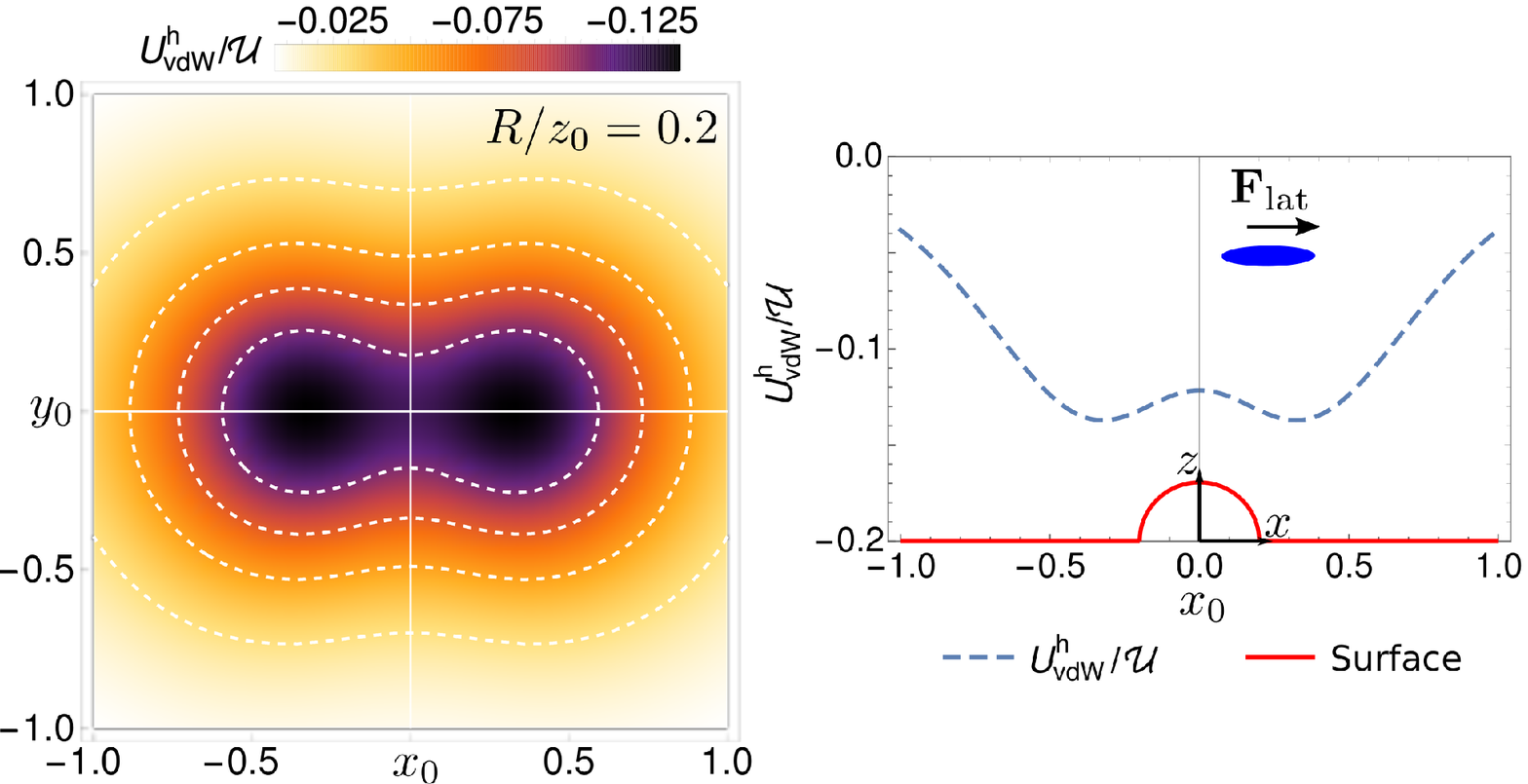, width=1 \linewidth}}
\caption{ 
Behavior of $U^{(\text{h})}_{\text{vdW}}/{\cal{U}}$, with ${\cal U}=\langle d_{p}^{2}\rangle/(64\pi\epsilon_{0}z_{0}^{3})$, versus ${x_0}/{z_0}$ and ${y_0}/{z_0}$, for a particle fixed at $z = z_0$.
In each panel, we consider this particle characterized by $\beta=0.2$ and oriented such that $\theta=\pi/2$ and $\gamma=0$ (its symmetry axis is along the $x$-direction).
From panel (a) to (b), we change the value of the ratio $R/z_0$ from $R/z_0 = 0.6$ to $R/z_0 = 0.2$.
The insets on the right show the behavior of $U^{(\text{h})}_{\text{vdW}}/{\cal{U}}$ with respect to $x_0$ $(y_0=0)$, and illustrate that the change in the behavior of the energy results in a sign inversion of the lateral vdW force (represented by the arrows).
}
\label{fig:energia-xy}
\end{figure}

To investigate the behavior of $U^{(\text{h})}_{\text{vdW}}$, let us start considering a class of particles whose tensor $ \langle \hat{d}_i \hat{d}_j \rangle $ diagonalized, in cylindrical coordinates, is represented by the matrix
\begin{equation} \label{eq:didj}
\langle \hat{d}_i \hat{d}_j \rangle=\begin{pmatrix}
\langle \hat{d}_n^2 \rangle&0&0\\
0&\langle \hat{d}_n^2 \rangle&0\\
0&0&\langle \hat{d}_p^2 \rangle
\end{pmatrix},
\end{equation}
which represents cylindrically symmetric polarizable particles, and thus the subscript $p$ and $n$ refer to the directions parallel and normal to the symmetry axis of the particle (here, we consider $\langle \hat{d}_p^2 \rangle \geq \langle \hat{d}_n^2\rangle$).
For a general orientation of this particle, we can write
\begin{align}
\langle\hat{d}_{\rho}^{2}\rangle & =\langle\hat{d}_{p}^{2}\rangle\left[\beta+\left(1-\beta\right)\sin^{2}\left(\theta\right)\cos^{2}\left(\gamma-\phi_{0}\right)\right], \\
\langle\hat{d}_{\phi}^{2}\rangle & =\langle\hat{d}_{p}^{2}\rangle\left[\beta+\left(1-\beta\right)\sin^{2}\left(\theta\right)\sin^{2}\left(\gamma-\phi_{0}\right)\right], \\
\langle\hat{d}_{z}^{2}\rangle & =\langle\hat{d}_{p}^{2}\rangle\left[\beta+\left(1-\beta\right)\cos^{2}\left(\theta\right)\right], \\
\langle\hat{d}_{\rho}\hat{d}_{z}\rangle & =\langle\hat{d}_{p}^{2}\rangle\left[\frac{1-\beta}{2}\sin\left(2\theta\right)\cos\left(\gamma-\phi_{0}\right)\right],
\end{align}
where $\beta = \langle\hat{d}_{n}^{2}\rangle/\langle\hat{d}_{p}^{2}\rangle $ $(0 < \beta \leq 1)$, and the angles $\theta$ and $\gamma$ describe the particle orientation, as illustrated in Fig. \ref{fig:particula-cilindricas}.
As a first case, let us consider a particle, kept constrained to move on the plane $z = z_0$, characterized by $ \beta = 0.2 $ and oriented with $\theta=\pi/2$ and $\gamma=0$ (its symmetry axis is along the $x$-direction).
In Fig. \ref{fig:energia-xy}, we show the behavior of $U^{(\text{h})}_{\text{vdW}}$ for $R/z_0 = 0.6$ [Fig. \ref{fig:energia-xy-r06}] and $R/z_0 = 0.2$ [Fig. \ref{fig:energia-xy-r02}].
In Fig. \ref{fig:energia-xy-r06}, note that the minimum point of $U^{(\text{h})}_{\text{vdW}}$ is over the origin, which means that the lateral vdW force always leads the particle to the peak of the hemisphere.
On the other hand, in Fig. \ref{fig:energia-xy-r02}, we decrease the value of the ratio $R/z_0$ and one notes that $U^{(\text{h})}_{\text{vdW}}$ now has two minimum points, and none of them coincide with the peak of the hemisphere.
This change in the minimum points means that a particle slightly displaced from the origin can feel a sign inversion in the lateral force, in the sense that instead of being attracted towards the hemisphere, the particle can be moved away from it.
These examples show that such a sign inversion in the lateral vdW force, which is a nontrivial geometric effect of the presence of the hemisphere on the conducting plane, is regulated by the ratio $R/z_0$.
Moreover, the existence of this effect is now confirmed by means of an exact calculation, with no restriction on the size of the hemisphere or the particle position, differently from Ref. \cite{Nogueira-PRA-2022}.

\begin{figure}[h]
\centering  
\subfigure[\label{fig:regioes-gamma-0}]{\epsfig{file=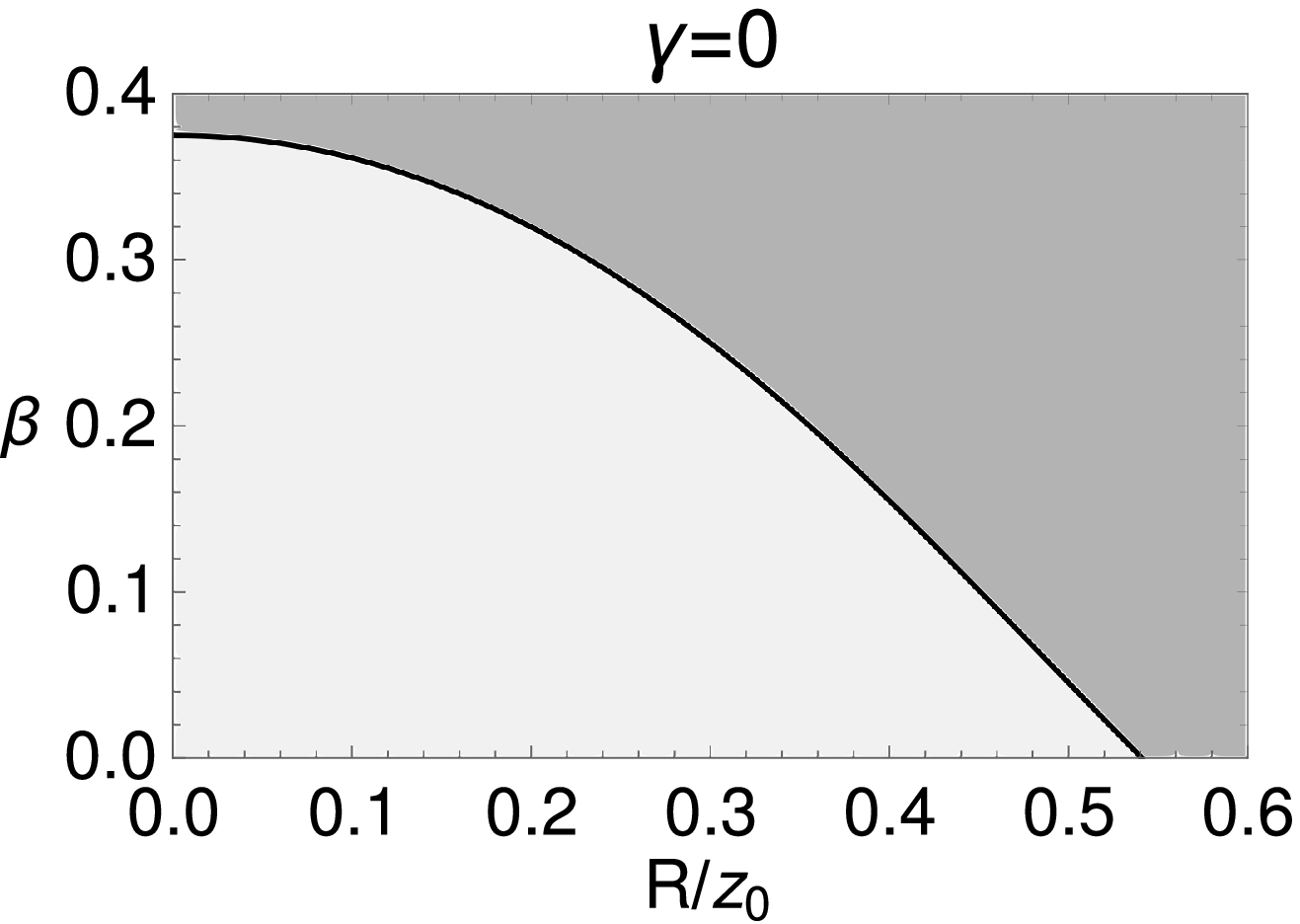, width=.48 \linewidth}}
\hspace{2mm}
\subfigure[\label{fig:regioes-gamma-30}]{\epsfig{file=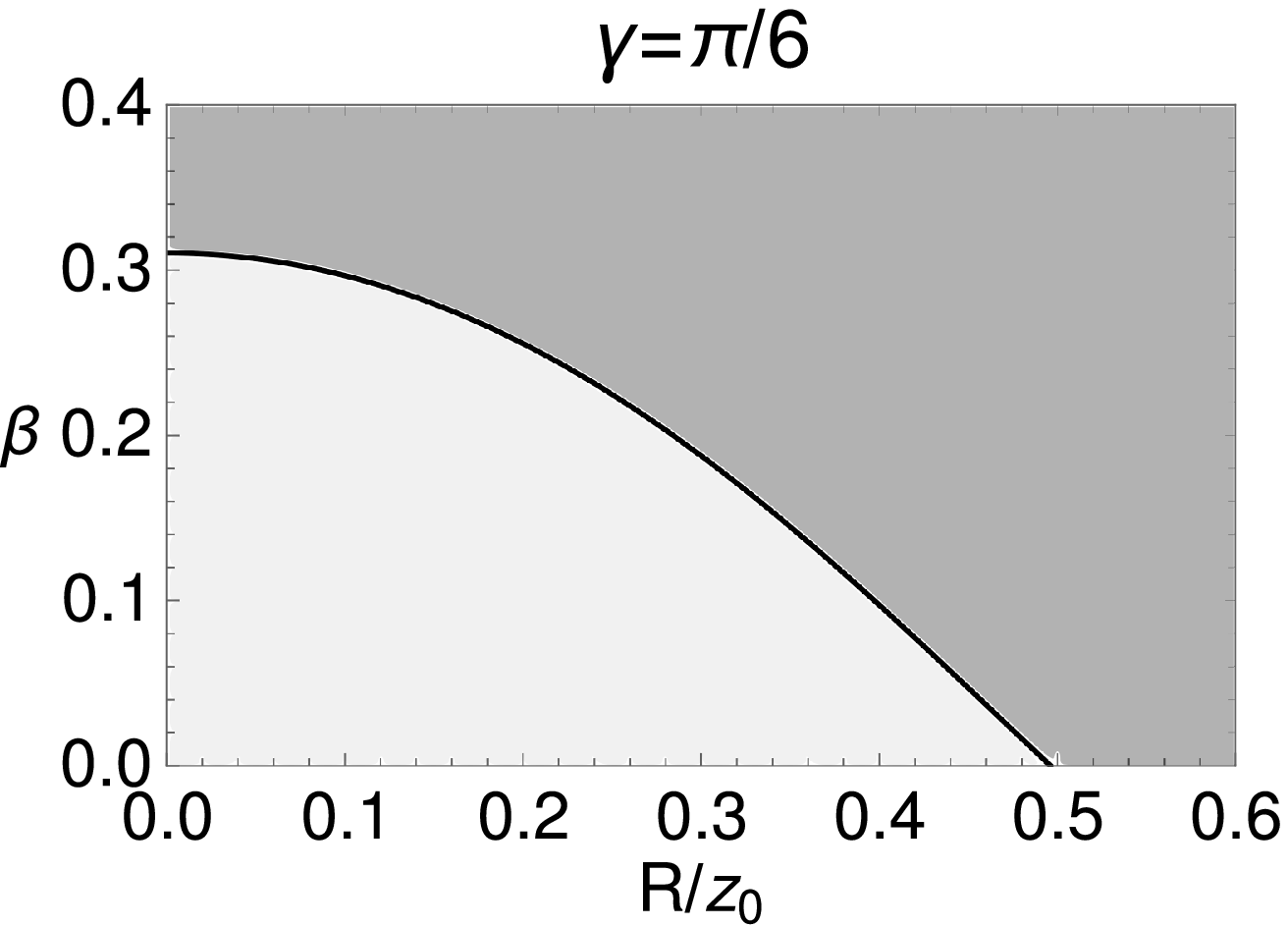, width=.48 \linewidth}}
\hspace{2mm}
\subfigure[\label{fig:regioes-gamma-60}]{\epsfig{file=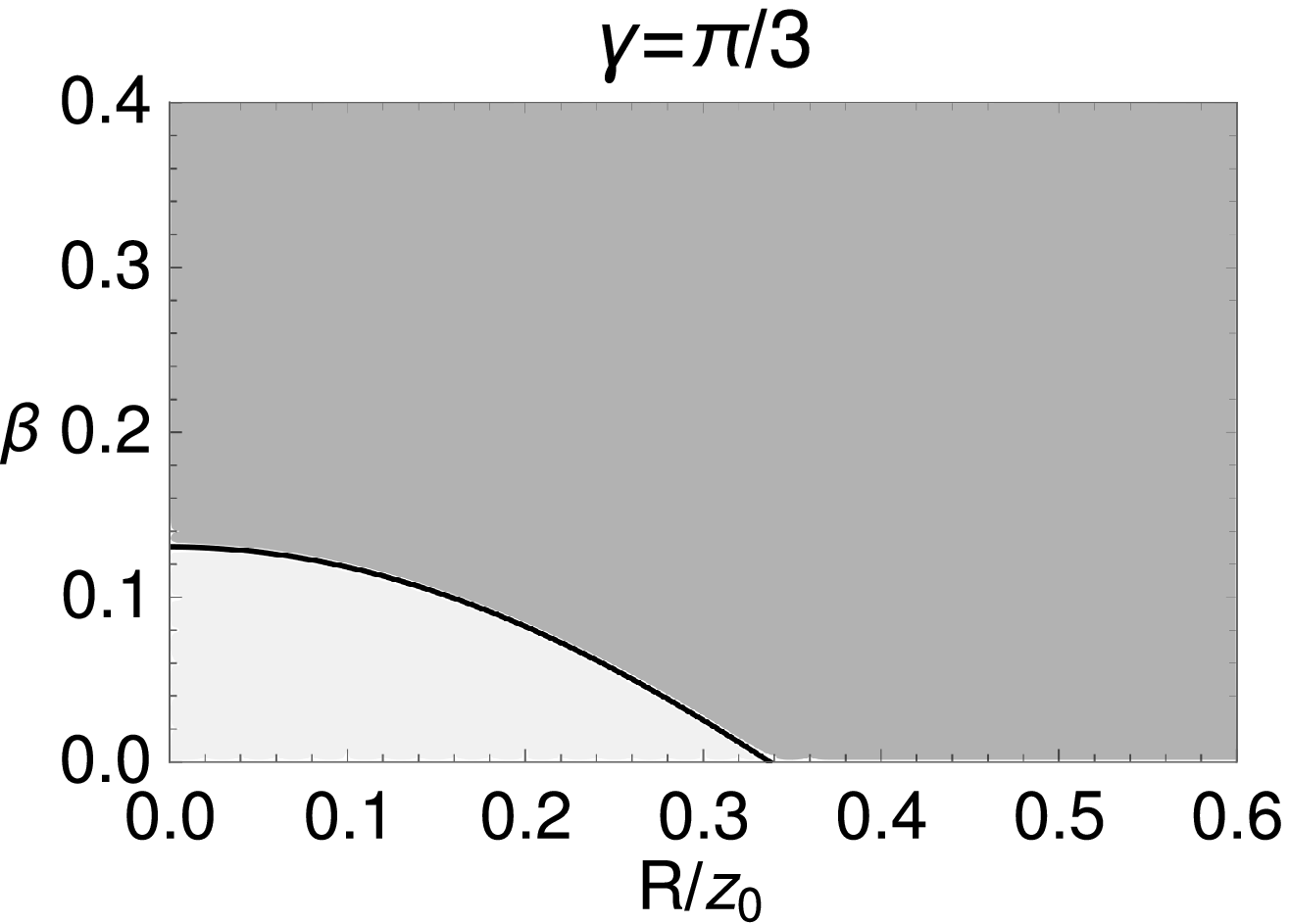, width=.48 \linewidth}}
\caption{ 
For a particle free to move along the $x$-axis $(y_0=0)$ and oriented with $\theta=\pi/2$, it is shown the configurations of $\beta$ and $R/z_0$ for which $x_0=0$ is a minimum (dark region) or a maximum (light region) point of $U^{(\text{h})}_{\text{vdW}}$.
These configurations are shown for (a) $\gamma=0$, (b) $\gamma=\pi/6$ and (c) $\gamma=\pi/3$.
We remark that, we just have a dark region when $\beta > 3/8$ in (a), $\beta > 9/29$ in (b), and $\beta > 3/23$ in (c).
}
\label{fig:regioes-gamma}
\end{figure}

\begin{figure}[h]
\centering
\subfigure[\label{fig:energia-orientacao-60}]{\epsfig{file=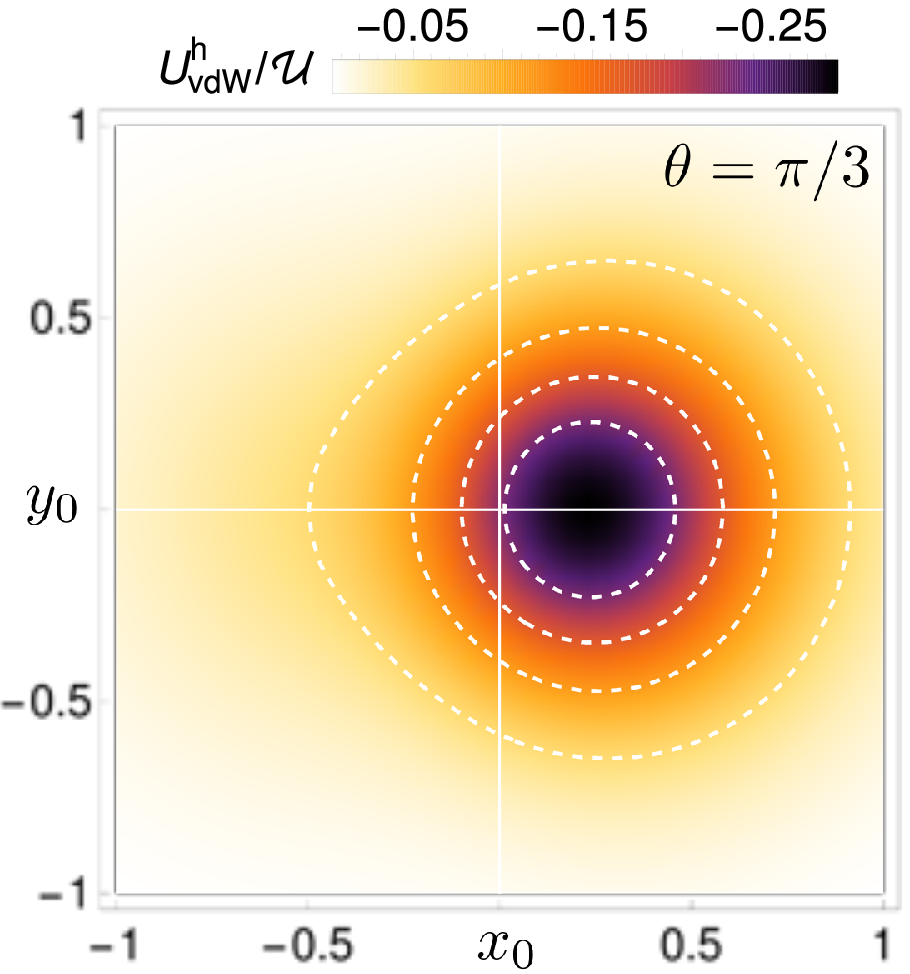, width=0.48 \linewidth}}
\hspace{2mm}
\subfigure[\label{fig:energia-orientacao-30}]{\epsfig{file=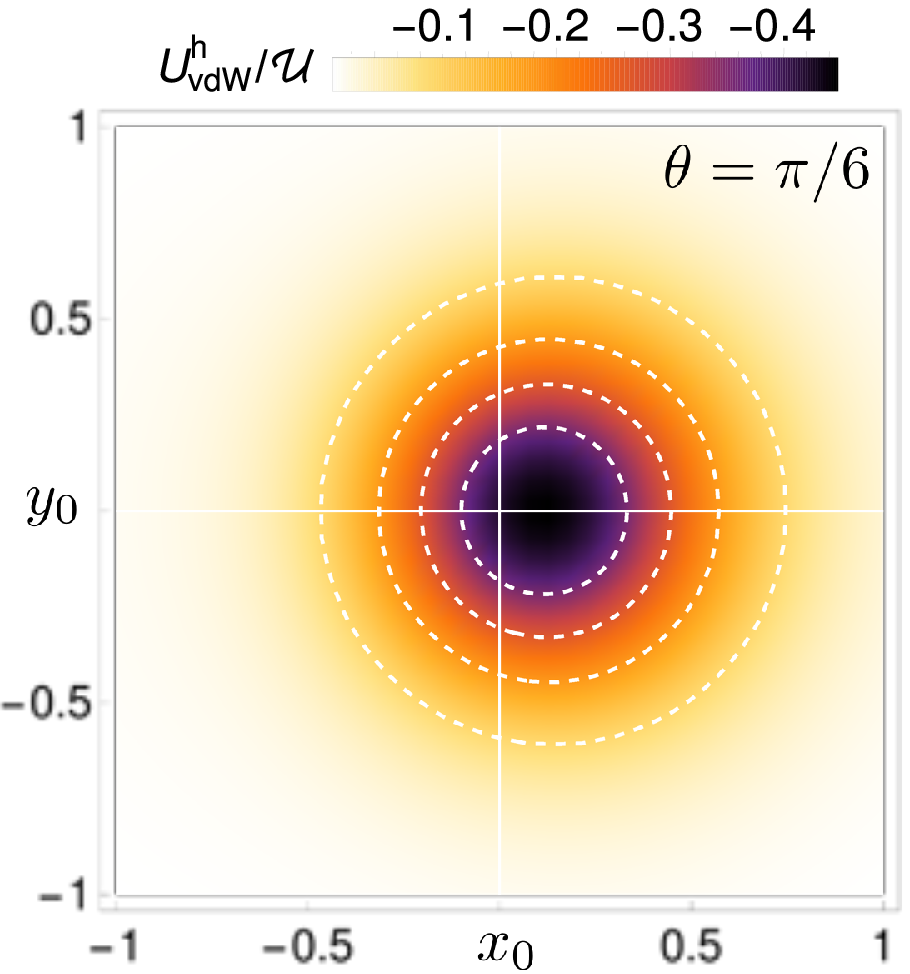, width=0.48 \linewidth}}
\hspace{2mm}
\subfigure[\label{fig:energia-orientacao-0}]{\epsfig{file=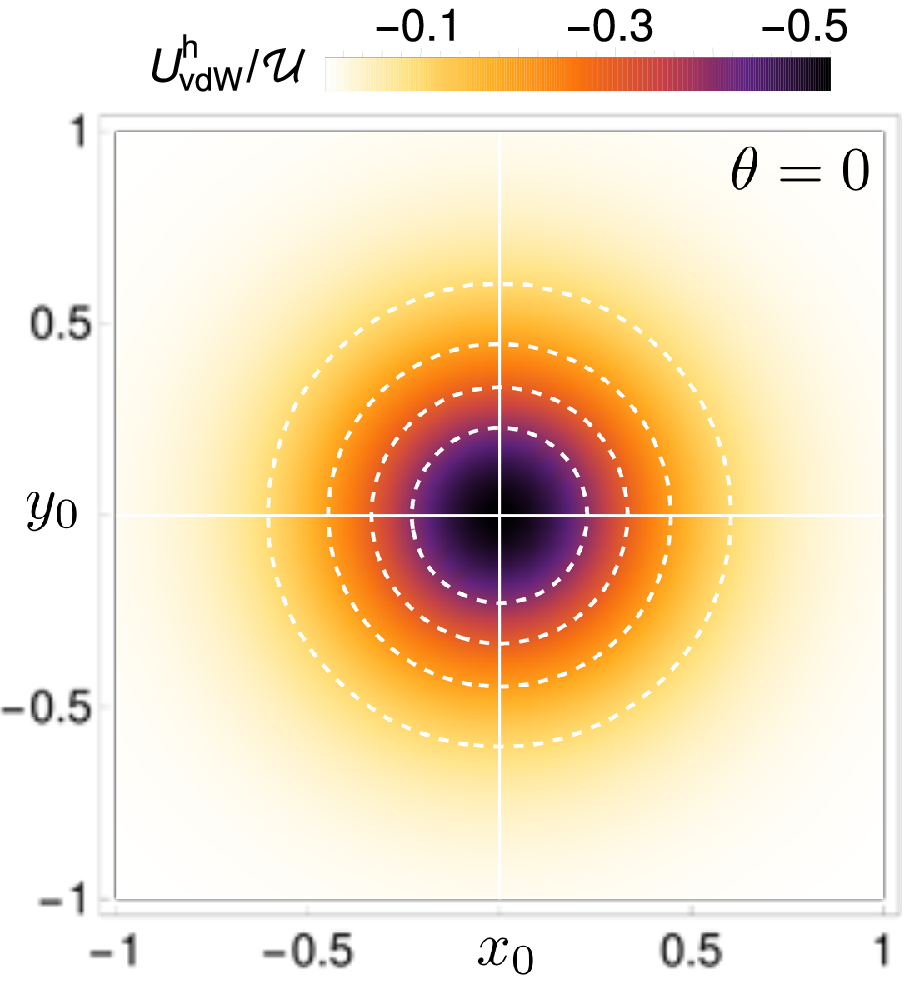, width=0.48 \linewidth}}
\caption{ 
Behavior of $U^{(\text{h})}_{\text{vdW}}/{\cal{U}}$, with ${\cal U}=\langle d_{p}^{2}\rangle/(64\pi\epsilon_{0}z_{0}^{3})$, versus ${x_0}/{z_0}$ and ${y_0}/{z_0}$, for a particle fixed at $z = z_0$.
In each panel, we consider $R/z_0=0.2$, and the particle characterized by $\beta=0.2$ and oriented with $\gamma=0$ and different values of $\theta$ (its symmetry axis is along the $xz$-plane).
We consider $\theta=\pi/3$ in panel (a), $\theta=\pi/6$ in panel (b), and $\theta=0$ in panel (c).
}
\label{fig:energia-orientacao}
\end{figure}

The change in the minimum points by manipulating the ratio $R/z_0$, as shown in Fig. \ref{fig:energia-xy}, also occurs for other values of $\beta$.
But, we remark that above a certain value of $\beta$, $U^{(\text{h})}_{\text{vdW}}$ will always have only one minimum point over the peak of the hemisphere, independently on the ratio $R/z_0$.
This means that, the mentioned sign inversion in the lateral vdW force is affected by the particle anisotropy.
In this way, for a particle free to move along the $x$-axis $(y_0=0)$ and oriented with $\theta=\pi/2$ and $\gamma=0$, in Fig. \ref{fig:regioes-gamma-0} we show the configurations of $\beta$ and $R/z_0$, such that the peak of the hemisphere ($x_0 = 0$) is a minimum (dark region) or a maximum (light region) point of $U^{(\text{h})}_{\text{vdW}}$.
The sign inversion in the lateral force is obtained when a transition from the dark to the light region occurs, and, thus, one can note that for $\beta > 3/8$ there are no manipulation of the ratio $R/z_0$ that makes possible the sign inversion in the lateral force.
In addition, the regions shown in Fig. \ref{fig:regioes-gamma-0} change depending on the orientation of the particle in the $xy$ plane, which is given by the angle $\gamma$ (see Fig. \ref{fig:particula-cilindricas}), and thus, we show these regions for other values of $\gamma$ in Figs. \ref{fig:regioes-gamma-30} and \ref{fig:regioes-gamma-60}.
This means that, behaviors similar to that shown in Fig. \ref{fig:energia-xy} also occur for other values of $\gamma$, but with the change in the minimum points occurring for different configurations of $\beta$ and $R/z_0$.
This not occur when we reorient the particle by changing the angle $\theta$.
In this situation, we have an asymmetric behavior of $U^{(\text{h})}_{\text{vdW}}$ when $0<\theta<\pi/2$, as shown in Figs. \ref{fig:energia-orientacao-60} and \ref{fig:energia-orientacao-30}.
When $\theta=0$ (symmetry axis of the particle along the $z$-direction), the minimum point of $U^{(\text{h})}_{\text{vdW}}$ is always over the origin [see Fig. \ref{fig:energia-orientacao-0}], independent on the value of $ \beta$ or the ratio $R/z_0$.
All of this show us that the sign inversion in the lateral vdW force is also affected by the particle orientation.

\section{Final remarks}
\label{sec-final-remarks}

Using the well known solution for the Green's function for the problem of
a perfectly conducting plane with a hemispherical protuberance with radius $R$ [Eq. \eqref{eq:G-exato}],
we investigated the vdW interaction [Eqs. \eqref{eq:U-0}-\eqref{eq:rij-calota}] between an anisotropic particle interacting with a perfectly conducting plane containing a protuberance with the shape of a hemisphere. 
We verified, via an exact calculation, the sign inversion in the lateral vdW force, 
so that, instead of pointing to the protuberance [Fig. \ref{fig:energia-xy-r06}], in certain situations the lateral vdW force points to the opposite direction [Fig. \ref{fig:energia-xy-r02}].
In the literature, such prediction considered that the height of the protuberance was very small when compared to the distance between the particle and the plane \cite{Nogueira-PRA-2022}. 
In contrast, here we discussed such effect with no restriction on the particle position and on the size of the hemisphere, obtaining the lateral vdW force for any value of the ratio $R/z_0$. 
We investigated how such nontrivial geometric effect depends on this ratio, and how the particle orientation and anisotropy affect the sign inversion (Figs. \ref{fig:regioes-gamma} and \ref{fig:energia-orientacao}).

The investigation of such nontrivial geometric effects on the lateral vdW force, presented here, contributes to a 
better understanding and control of the interaction between a neutral anisotropic polarizable particle and a non-planar surface.

\begin{acknowledgments}
The authors thank Alexandre P. da Costa for careful reading of this paper an fruitful discussions.
L.Q. and E.C.M.N. were supported by the Coordena\c{c}\~{a}o de Aperfei\c{c}oamento de Pessoal de N\'{i}vel Superior - Brasil (CAPES), Finance Code 001.
\end{acknowledgments}
%


%

\end{document}